\documentclass[twocolumn,showpacs,amsmath,graphicx,amssymb]{revtex4}

\usepackage{xcolor}
\usepackage{amsfonts}
\usepackage{amssymb}
\usepackage{amsmath}
\usepackage{graphicx}

\makeatletter

\let\saved@includegraphics\includegraphics
\AtBeginDocument{\let\includegraphics\saved@includegraphics}
\renewenvironment*{figure}{\@float{figure}}{\end@float}
\makeatother

\begin{document}

\title{Quantum materials interfaces: graphene/Bismuth (111) heterostructures}
\author{Ivan I. Naumov  and Pratibha Dev  }
\affiliation{Department of Physics and Astronomy, Howard University, 2355 6th Street, NW, Washington, D.C. 20059, USA}
\date{\today}
\begin{abstract}

Heterostructures involving graphene and bismuth, with their ability to absorb light over a very wide energy range, are of interest for engineering next-generation opto-electronics. Critical to the technological application of such heterostructures is an understanding of the underlying physics governing their properties. Here, using first-principles calculations, we study the interfacial interactions between graphene and bismuth thin-films. Our study reveals non-intuitive phenomena associated with the moir\'e-physics of these superlattices. We show a preservation of graphene-derived Dirac cones in spite of proximity to a substrate with large spin-orbit coupling, a greater influence of graphene on the electronic structure properties of bismuth, and the surprising presence of a magnetic solution, only slightly higher in energy (by several meV) than the non-magnetic structure, possibly validating experiments. Such subtle and unanticipated phenomena associated with the moir\'e-physics are expected to play key roles in the practical applications of heterogeneous assemblies of two-dimensional quantum systems.

\end{abstract}

 \maketitle
\section{Introduction}
 Low-dimensional heterosructures made by bringing together different (or even the same, but
mis-orientated) two-dimensional (2D) materials form moir\'e-superlattices. These moir\'e-superlattices often display new and unusual properties, providing opportunities for fundamental physics and applications~\cite{novoselov2}. Some notable emergent phenomena due to vertical stacking of 2D materials are: (a) flat bands and superconductivity observed in bilayer graphene with small twist angles (magic angles) between the two layers~\cite{cao}, and (b) quantum anomalous Hall effect  \cite{ren2}, spin-orbit torque~\cite{dyrdal}, transport magneto-anisotropies~\cite{lee2} in a graphene monolayer with substrate-induced exchange and spin-orbit couplings.

In the last few years, a considerable interest has been expressed in heterostructures involving graphene and bismuth (Bi). These composite structures are remarkable for many reasons, including enhanced light absorption over a very wide energy spectrum, encompassing the  infrared, visible and ultraviolet frequency ranges.  It was demonstrated, for example, that bismuth nanoparticles integrated with graphene can serve as an effective photocatalyst. Such a heterostructure is stable and displays high photocatalytic activity across the whole solar spectrum \cite{yan}. The ability to absorb light over a wide range of frequencies also makes graphene/Bi interfaces ideal for use in photodetectors. In this context, a particular composite structure -- bismuth nanowire arrays capped by graphene -- was found to be especially promising as it exhibited an enhanced, fast and broadband photoresponse \cite{jin, huber}. 

To date, no prior theoretical studies have examined graphene-Bi heterostructures, although there are several studies focused on a related problem of how the adsorbed or intercalated Bi atoms interact with graphene~\cite {akturk, hsu, warmuth}. These theoretical works, although instructive, cannot be used to understand and explain the photoresponse of graphene-bismuth interfaces that were investigated in the aforementioned experiments. Motivated by the lack of an understanding of the physics of the system at the atomistic level and the experimental works on this heterostructure~\cite{jin, huber}, here  we study  graphene/Bi interface. We address several unresolved questions: (a) What is the most stable crystal structure of the interface? (b) Does the Dirac electron structure remain intact in the heterostructure, and why? (c) What is the charge transfer between the constituents? (d) How strong is the spin-orbit or Rashba splitting of the surface states? In the process, we shed light on how heterogenous assemblies of quantum materials can result in unexpected and sometimes subtle, yet important interfacial effects.  One of the unexpected results is the preservation of graphene-derived Dirac cones, albeit with a small gap at the Dirac points due to breaking of the sub-lattice symmetry. This is a surprising result as graphene's electronic-structure properties are expected to be greatly influenced due to proximity to the Bi-substrate, with Bi being a very heavy element with a large spin-orbit coupling (SOC). On the other hand, there is a greater influence of graphene on the electronic structure properties of Bi substrate, with the Bi-derived surface states showing an atypical Rashba-type effect. Another surprising result is the presence of a magnetic solution, only slightly higher in energy (by several meV) than the non-magnetic structure, giving a possible explanation to experimental observations \cite {takayama1, takayama2}, which have been at variance with theoretical results thus far. In the magnetic structure, the moments are mostly contributed by the surface/sub-surface bismuth atoms. The broken time-reversal symmetry in the magnetic structure lifts the degeneracy of the Bi-derived surface states at the time-reversal invariant momenta (TRIM), as is observed in experiments.  These phenomena associated with the moir\'e physics are expected to play key roles in the practical applications of these materials.

In addition, our work helps to shed light on the experimentally-observed enhanced photoresponse in the heterostructure formed between single layer graphene on Bi nanowires.  This phenomenon was interpreted differently in Refs.~\cite{jin} and \cite {huber}. According to Jin \textit{et al.}, the major player here is graphene, and its improved light absorption is due to the opening of a band gap. This gap, in turn, is assumed to be caused by the wrinkled surface and anisotropic tensile strain~\cite{jin}. Huber \textit{et al.}, however, argued that the generated photocurrent is mostly of photovoltaic nature, \textit{i.e.}, based on the separation of generated electron-hole (\textit{e-h}) pairs by the built-in electric field across the interface (it was assumed that charge transfers from Bi to graphene). In this mechanism, in contrast to that suggested in Ref. \cite{jin}, both of the constituents of the system play crucial roles:  while graphene serves as a channel material, Bi nanowires absorb light due to a higher density of states (DOS) at the Fermi level. As will be discussed in subsequent sections, our results support the assumption made by Huber \textit{et al.}, that the zero-bias photocurrent generated on the graphene/bismuth interface is due to the interfacial electric field.

\section{Methods}

We performed first-principle calculations using the projector-augmented wave method~\cite{kresse1} as implemented in the Vienna ab-initio simulation package (VASP)~\cite{kresse2}. Considering that the diameter of nanowires used in the experiments~\cite{jin, huber} is large  (200-250\,nm), their interaction with a graphene layer can be understood by studying an interface between graphene and a bismuth thin-film.  In our study, we used a Bi (111) surface as it is known to be the most important Bi surface for practical applications~\cite{hofmann}. The (111)-terminated Bi surface was simulated by thin Bi-films formed by stacking several (from 1 to 8)  Bi (111) bilayers (BL) along the trigonal axis, each BL representing a buckled honeycomb lattice. Results are presented for the composites with a 6-BL-thick Bi(111) substrate in the short-period commensurate structures, while a 3-BL-thick Bi(111) substrate is used for the long-period commensurate structures. We should note that for larger structures considered in this work, one can also use a single BL Bi (111) as a  substrate, to ease the computation burden. Though this approximation overestimates the quantum confinement of  the substrate, it nevertheless captures the main features of the interface interaction between graphene and Bi (111) surface. 

We adopted the PBE generalized gradient approximation (GGA) to describe the exchange-correlation potential  \cite{perdew}. The spin-orbit coupling (SOC) which plays a significant role in the electronic structure of Bi, was taken into account in the self-consistent calculations of energy bands. 
The kinetic energy cutoff was set to 500\,eV. During structural relaxations the in-plane lattice parameter of the Bi (111) film was fixed to the bulk value and atoms were allowed to move until the difference in total energies between two ionic steps was below $1\times 10^{-4}$\,eV.  A 25\,{\AA}-thick vacuum layer was chosen to prevent interaction between the replica slabs. Our calculations included dipole corrections to avoid interactions of the heterostructure with its images induced by the asymmetric graphene-Bi composite with a non-vanishing surface dipole density. In order to take van der Waals interactions into account, we adopted the many-body dispersion energy method~\cite{tkachenko1} since it leads to interlayer distances close to the experimental values, both in graphite and bulk bismuth.  A 16$\times$16$\times$1 Monkhorst-Pack \emph{k}-point grid has been used in the case of an ideal graphene lattice containing only two atoms per cell. Approximately the same \emph{k}-point density was kept in going from the 2D honeycomb lattice to more complicated structures. 

Since the actual structure of graphene on a Bi (111) substrate is experimentally undetermined and can be incommensurate, we investigated a large number of crystal approximants to the incommensurate interfaces. The next section details the geometrical and symmetry-related aspects of the possible graphene/Bi (111) heterostructures considered in this work.

\section{ possible commensurate graphene/Bi(111) heterostructures}
In this section, we consider some purely geometric aspects of commensurate hexagonal hetrostructures and highlight symmetry-based aspects that are needed for future discussions.

In what follows, we assume that a graphene sheet is placed on a Bi (111) film that has hexagonally arranged surface atoms. We want to create a set of supercells or moir\'e  patterns
 that are commensurate with both the graphene substrate and film.  Let
 $\boldsymbol{a}_{1}=\frac{a_{g}}{2}\left(\sqrt{3}%
,1\right)$ and  $\boldsymbol{a}_{2}=\frac{a_{g}}{2}\left( -\sqrt{3},1\right) $
be the primitive lattice vectors of the free-standing graphene sheet.  Similarly, we can select the in-plane primitive lattice vectors
of a Bi (111) film as $\boldsymbol{b}_{1}=\frac{a_{Bi}}{2}\left( \sqrt{3}%
,1\right)$ and  $\boldsymbol{b}_{2}=\frac{a_{Bi}}
{2}\left( -\sqrt{3},1\right) $; here, ${a_{g}}$ and ${a_{Bi}}$ are the corresponding lattice constants.
Possible hexagonal supercells commensurate with the graphene lattice are characterized by the lattice vectors:
\begin{equation}
{\binom{\boldsymbol{t}_{1}}{\boldsymbol{t}_{2}}}=\left(
\begin{array}{cc}
\; n & m\\
-m &\, n-m%
\end{array}%
\right) {\binom{\boldsymbol{a}_{1}}{\boldsymbol{a}_{2}},}  \label{eq:t1t2}
\end{equation}

\noindent where $n$ and $m$ are integers (both negative and positive), and $|\boldsymbol{t}_{1}|=|\boldsymbol{t}_{2}|=|t|=a_g \sqrt{n^{2}-nm+m^{2}}$.
In general, the vectors $\boldsymbol{t}_{i}$  and $\boldsymbol{a}_{i}$  are not parallel to each other and form an angle:
\begin{equation}
\theta_{g} =\arccos \left(\frac{n-m/2}{\sqrt{n^2-nm+m^2}}\right),  \label{eq:3}
\end{equation}

Using Woods notation one can label the commensurate supercells as $\left(f(n,m)\times f(n,m)\right)$R$
\theta_{g}$, where the symbol R means rotation by a corresponding angle and $f(n,m)=\sqrt{n^{2}-nm+m^{2}}$. As
in the case of graphene, possible hexagonal supercells commensurate with the Bi substrate have lattice
vectors given by:

\begin{equation}
{\binom{\boldsymbol{T}_{1}}{\boldsymbol{T}_{2}}}=\left(
\begin{array}{cc}
\; N & M\\
-M &\, N-M%
\end{array}%
\right) {\binom{\boldsymbol{b}_{1}}{\boldsymbol{b}_{2}},}  \label{eq:T1T2}
\end{equation}

\noindent where, $|\boldsymbol{T}_{1}|=|\boldsymbol{T}_{2}|=|T|=a_{Bi} \sqrt{N^{2}-NM+M^{2}}$. To be  mutually commensurate, the graphene and  Bi (111) supercells must have identical lattice parameters:   $\boldsymbol{t}_{1} = \boldsymbol{T}_{1}$ and $\boldsymbol{t}_{2} = \boldsymbol{T}_{2}$. This condition is equivalent to demanding that the graphene and Bi superlattices obey the so-called diophantine equation: $a_g\times f(n,m)=
a_{Bi}\times f(N,M)$. The twist angle between the two lattices is given by:
\begin{equation}
\theta=\theta_{g} -\theta_{Bi},  \label{eq:4}
\end{equation}

Since Bi is a substrate, its in-plane lattice parameter $a_{Bi}$ should be fixed to a bulk value. This condition forces graphene to accept a lattice parameter $a_g = a_{Bi}\times f(N,M)/f(n,m)$, which is in general different from that in free-standing graphene $a_g ^0$. In Table~\ref{tab:table1} we present the four smallest
crystal approximants to the incommensurate interface;  both the $(n,m)-(N,M)$ indexing and Woods nomenclature are used. One can see that in the first approximant the graphene lattice parameter $a_g$ is larger, in the second smaller, in the third again larger, and in the fourth is almost exactly equal to a freestanding  value, $a_g ^0$ (4.533\,{\AA}).

\begin{table}[tbp]
\caption{Four smallest  crystal approximants or moir\'e supercells to the incommensurate interface between
graphene and Bi(111) film.  Note that for all the  Bi-supercells, $N\neq 0$ but $M=0$, so that $\theta_{Bi}$=0 and $\theta =\theta_{g}$.}
 \label{tab:table1}
\begin{ruledtabular}
  \begin{tabular}{|c|c|c|}
  Structure &moir\'e supercell &$ a_g /a_g ^0$\\
  \tableline
   I & \begin{tabular}{@{}c@{}} $(n,m)=(1,2), (N,M)=(1,0)$\\ $(\sqrt{3}\times \sqrt{3})R30^0$  Gr on $(1\times 1)$ Bi\end{tabular} &1.060\\
 \tableline
  II &\begin{tabular}{@{}c@{}} $(n,m)=(2,0), (N,M)=(1,0)$\\ $ (2\times 2)$  Gr on $(1\times 1)$ Bi\end{tabular} &0.919\\
  \tableline
  III&\begin{tabular}{@{}c@{}} $(n,m)=(1,4), (N,M)=(2,0)$\\ $(\sqrt{13} \times \sqrt{13})R13.9^0$  Gr on $(2\times 2)$ Bi\end{tabular}&1.018 \\
 \tableline
 IV &\begin{tabular}{@{}c@{}}  $(n,m)=(1,6), (N,M)=(3,0)$\\ $(\sqrt{31} \times \sqrt{31})R8.9^0 $  Gr on $(3\times 3)$ Bi\end{tabular}&1.000 \\

     \end{tabular}
   \end{ruledtabular}
\end{table}

The described relations  between $\boldsymbol{t}_{1,2}$ and  $\boldsymbol{a}_{1,2}$, $\boldsymbol{T}_{1,2}$ and
$\boldsymbol{b}_{1,2}$, as well as identity of $\boldsymbol{t}_{i}$ with  $\boldsymbol{T}_{i}$ establish a one-to-one
correspondence between the moir\'e, graphene and Bi (111) Brillouin zones (BZ). Using simple algebra one can show that the $\boldsymbol{K}_g$ and $\boldsymbol{K}^\prime_g$   momentum of graphene, for example,
 are folded to a $\boldsymbol{\Gamma}$ moir\'e point if $(n+m)$ is a multiple of 3.  If, however, ($n+m$) is not a multiple of 3,
then  the momenta  $\boldsymbol{K}_g$ and $\boldsymbol{K}^\prime_g$ map to different  moir\'e  $k$  points:
$\boldsymbol{K}_g \to \boldsymbol{K},\boldsymbol{K}^\prime_g \to \boldsymbol{K}^\prime $ or  $\boldsymbol{K}_g
\to \boldsymbol{K} ^\prime $, $\boldsymbol{K}^\prime_g \to \boldsymbol{K}$. Further, if both $n$ and $m$ are multiples of 2,
then the $\boldsymbol{M}^\prime_g$ points are translated into $\boldsymbol{\Gamma}$. If only $n$ (or only $m$), or none of them is a multiple of 2, then
 $\boldsymbol{M}_g  \to \boldsymbol{M}$. The same relations are applicable to  Bi (111) thin films, if  the integers  $N$ and $M$ are considered
instead of  $n$ and $m$.

It should be stressed that the two pairs of integers,  $(n,m)$ and $(N,M)$,  and the  twist angle $\theta$  determine only the translation symmetry  of the moir\'e pattern. To define the point  symmetry we need to know the position of  the axis of twisting. When this axis coincides with the trigonal ($C_3$) axis of Bi (111) and passes through the carbon atom or hexagon center of graphene, the point group  of the whole heterostructure is also $C_3$. If the twisting axis does not coincide with the trigonal axis and/or does not pass through a carbon atom or hexagon, the point group reduces to $C_1$. In any case, the presence of the Bi (111) substrate breaks the symmetry between the $A$ and $B$ graphene sublattices, and will open a gap at the Dirac points. The splitting of the Dirac states can be additionally  facilitated in some moir\'e structures  due to mixing of electronic states belonging to two different valleys. In such structures, the graphene $\boldsymbol{K}_g$ and $\boldsymbol{K}^\prime_g$ points map to the $\boldsymbol{\Gamma}$  moir\'e point; they are exemplified by structure-I from Table~\ref{tab:table1}.

\section{Atomic and electronic properties of heterostructures}
In experiments, the composite structure may sample all or most of the registries and different strain levels listed in Table~\ref{tab:table1}. Hence, in the following sections, we consider all of the moir\'e supercells. Even so, particular attention has been given to two smallest supercells, I and II, because they are simple and reflect many features of larger nearly stress-free structures. To find the most stable atomic configuration for each structure, we not only relaxed the atomic positions but also considered different horizontal shifts of the graphene with respect to the Bi thin-film.  We found that in the most stable configurations, graphene's hexagons tend to be centered above the Bi atoms in the topmost layer. The barrier for the relative shifting is found to be rather small. It ranges from 5.0 to 7.5\,meV/atom, which is typical for van der Waals materials. The distance between graphene layer and substrate was found to be 3.40\,{\AA}, with a variation of about $\pm0.10$\,{\AA} for different commensurabilities. 

\subsection  {Structure-I: $(\sqrt{3}\times \sqrt{3})$  graphene  on $(1\times 1)$ Bi}
\begin{figure*}[tpbh] 
\begin{center}
\includegraphics[width=0.8\textwidth]{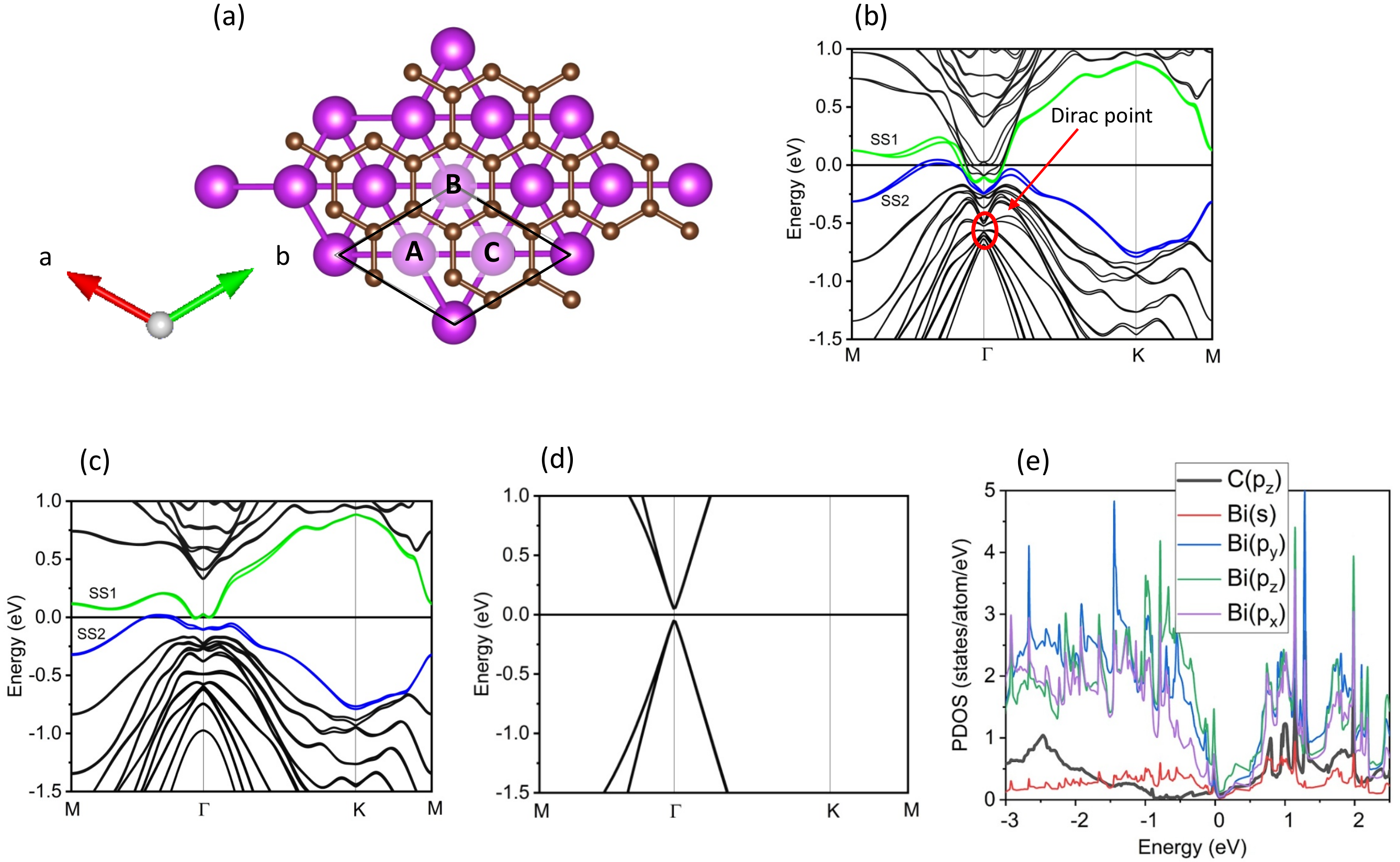}
\end{center}
 \vspace{-12pt}
\caption{(a)   A top view of the moir\'e supercell $(\sqrt{3}\times \sqrt{3})R30^0$  Gr on $(1\times 1)$  Bi (111).  Brown  (the top layer) and pink (the lower layers)  colors represent  C and Bi atoms,
respectively. The solid black lines enclose an in-plane unit cell of the composite.  The carbon hexagons have a C$_3$ symmetry and centered  on the  Bi-atoms marked as A, B, and C, from  S,  (S--1) and (S--2) surface layers, respectively.
 (b)  Band structure along the high symmetry lines, using E$_F$ as the reference energy. (c) Band structure for a model system of a Bi thin-film, created by removing graphene from the heterostructure. (d) Band structure for another model system of freestanding graphene, created by removing the Bi layers from the composite structure. (e) The projected density of states (PDOS) corresponding to the composite structure, showing negligible hybridization between the carbon-derived and Bi-derived states within 0.5 eV around E$_F$.}
\label{fig:fig1}
\end{figure*}


For this smallest supercell, $n+m$ is multiple of 3 and therefore,  the two Dirac cones associated with  graphene are translated into the $\overline{\Gamma}$ point. They should be gapped due to the inter-valley scattering. There is, however, an additional factor that contributed to the gap at the Dirac points --- the broken symmetry between the two graphene sublattices. Indeed, the lowest-energy structure for this superlattice is the one in which graphene's hexagons are centered on the Bi atoms of the top, second or third layer [Fig.~\ref{fig:fig1}(a)].  Thus, all of the initially equivalent carbon hexagons now belong to three different groups depending on which Bi-atom they surround. The hexagons from each group have a C$_{3}$ symmetry with slightly different alternating bonds. As the hexagons from different groups always share a common side, in going from one group to another the alternating bonds change as ($\alpha, \beta) \to (\alpha, \gamma)$ or ($\alpha, \beta) \to (\gamma,  \beta)$, where $\alpha$, $\beta$ and $\gamma$ are the three different calculated bond lengths. We thus see that the $A$ and $B$ graphene-site symmetry is broken.

 Our first-principles calculations confirm that the graphene-derived Dirac cone indeed shows up in the vicinity of $\overline{\Gamma}$ [Fig.~\ref{fig:fig1}(b)]. As expected, a sizable gap of 70\,meV is formed between the upper and lower part of the Dirac cone. This gap is much larger than the gap of 0.024\,meV in pristine graphene, which is caused by intrinsic spin-orbit coupling (SOC)~\cite{abdelouahed,gmitra}. Nevertheless, one might suspect that along with the combined effects of the two factors mentioned in the previous paragraph, there may be enhanced SOC within graphene induced by its proximity to a heavy-element substrate with large SOC. This will be shown not to be the case towards the end of this sub-section. Also, note that the cone is downshifted by $\sim0.55$\,eV, implying strong $n$-doping of the graphene layer.  

 Along the $\overline{\Gamma}$$\overline{M}$ direction, except the region close to $\overline{\Gamma}$, the calculated bands are similar to that found in pristine Bi (111) ultra thin-films~\cite{koroteev, hirahara3, ito1,chang}. Among them, there are two surface bands, SS1 and SS2, shown in green and blue respectively in Fig.~\ref{fig:fig1}(b). In pristine Bi (111) thin films, SS1 and SS2 are spin degenerate and each of them can be viewed as a hybrid of two surface states localized on opposite (top and bottom) sides and having opposite spins. In our case of the graphene/Bi (111) heterostructure, both SS1 and SS2 surface-bands are slightly spin-split, especially between the $\overline{\Gamma}$-  and $\overline{M}$-points [Fig. ~\ref{fig:fig1}(b)]. This spin-splitting will be discussed more fully in the next subsection. Here, however, we want to explicitly demonstrate that it is caused by the presence of the graphene layer which breaks the inversion symmetry of the heterostructure as a whole and lifts the Kramers degeneracy everywhere inside the BZ. Indeed, when the graphene layer is removed from the top of the Bi substrate while keeping the Bi atoms fixed, the splitting along $\overline{\Gamma}-\overline{M}$ practically disappears [Fig.~\ref{fig:fig1}(c)]. A weak residual splitting is present even after removing of the graphene layer, which can be traced to the fact that the mirror symmetry in the Bi thin-film (with respect to the surface-parallel plane passing through the center) still remains broken due to frozen graphene-induced lattice distortions. 

\begin{figure*}[tpbh] 
\begin{center}
\includegraphics[width=0.85\textwidth]{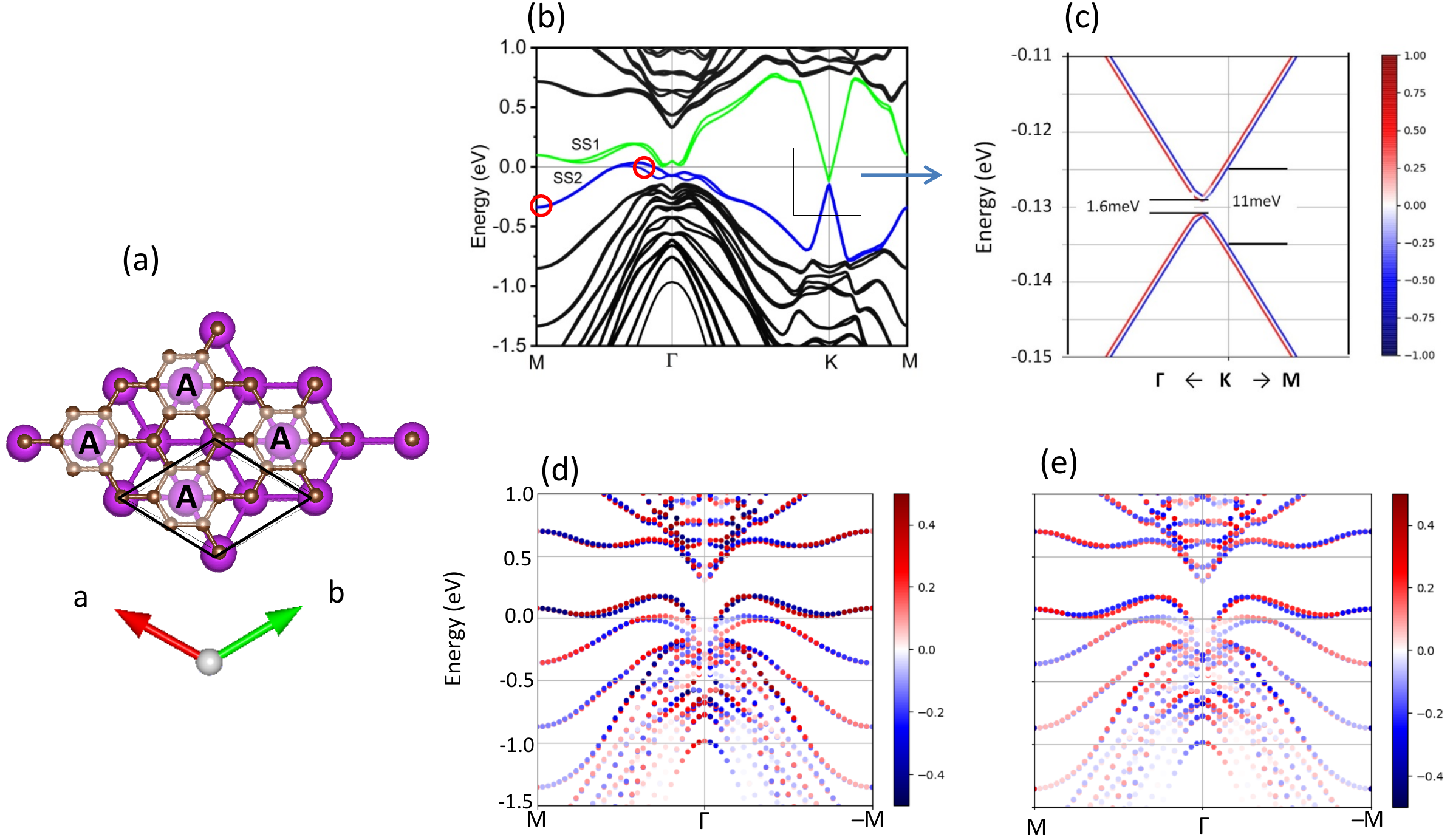}
\end{center}
\vspace{-12pt}
\caption{(a)   A top view of the moir\'e supercell $( 2\times 2)$ Gr on  $(1\times 1)$ Bi (111).  The solid black lines enclose an in-plane unit cell of the supercell. The topmost Bi atoms are labeled A.  (b)  Band structure along the high symmetry lines. (c)  A zoom into the region framed by black rectangle and containing the Dirac point in 
the panel (b), with the resolved spin projections  perpendicular to the  
 \emph{k}  vector. Colors code the expectation  values  of the projections. (d,e). Spin-resolved band structure from  $\overline{\Gamma}$ to  $\overline{M}$  and from $\overline{\Gamma}$ to --$\overline{M}$, with the $\pm S_x$ and  $\pm S_y$ spin projections, respectively.  Note that the $\pm S_z$ projections are relatively small.}.
 \label{fig:fig2}
\end{figure*}

If we remove the Bi-film instead of the graphene layer, then the heterostructure transforms into another model system--a freestanding graphene sheet with frozen Bi-induced lattice distortions. The band structure of this system is characterized by a band gap which is very close to that found in the heterostructure itself  -- 70\,meV [Fig.~\ref{fig:fig1}(d)]. This suggests that the gap in the Dirac cones opens up mainly due to to breaking of A- and B-sublattice symmetry and the intervalley scattering, whereas the proximity-induced SOC plays a minor role in such an opening. Further, panels (b) and (d) in Fig.~\ref{fig:fig1}, imply that there is hardly any hybridization between the $\pi$-derived bands of graphene and  $sp$-derived states of Bi, despite the fact that they have similar energies. This becomes especially clear from the projected density of states shown in Fig.~\ref{fig:fig1}(e), which shows very weak hybridization between the carbon $p_z$ states and Bi $s$, $p_x$, $p_y$ and $p_z$ states within 0.5  eV around E$_F$. The reason is that they do not overlap much spatially as no carbon atoms sit immediately above the topmost Bi-atoms [Fig.~\ref{fig:fig1}(a)]. It should be noted, however, that the hybridization becomes much more pronounced for the energies above 0.5\,eV relative to E$_F$, as seen from  Fig.~\ref{fig:fig1}(e).  

\vspace{-10pt}

\subsection{Structure-II: $(2\times 2)$  graphene  on $(1\times 1)$ Bi}  
\vspace{-6pt}

Here, the lowest energy configuration corresponds to a structure in which some carbon hexagons are centered above the Bi atoms in the topmost layer [Fig.~\ref{fig:fig2}(a)]. The average distance between graphene layer and substrate is 3.37\,{\AA}. The point symmetry of this structure is $C_3$ because the trigonal axis passes through the carbon atom. Here, the momentum $\boldsymbol{K}_g$ and $\boldsymbol{K}^\prime_g$ of graphene are folded to the $\boldsymbol{K}^\prime$ and \emph{k} points of the heterostructure, respectively, since ($n+m$) is not multiple of 3.

Figure ~\ref{fig:fig2}(b) is a plot of the the band structure of the composite structure. As can be seen from the band structure plot, the bands along the $\overline{\Gamma}$$\overline{M}$ direction are Bi-derived, and they are graphene-derived in the vicinity of the $\overline{K}$-point. This essential feature stems from the fact that the graphene and substrate orbitals barely hybridize in the near-E$_{F}$ region because they are well separated in \emph{k}--space.

As in the case of structure-I, here we observe two surface states, SS1 and SS2, dispersing along  $\overline{\Gamma}$$\overline{M}$. Again, they are spin-split inside the interval (but not at points $\overline{\Gamma}$ and $\overline{M}$ themselves, which are TRIM points) due to structural asymmetry induced by the presence of the graphene layer. Though similar, this is not a typical Rashba-type effect because it involves the states from the opposite surfaces, rather than the same surface. Such a splitting at a given \emph{k}-point leads to a global spin polarization due to asymmetry between the local polarizations on the opposite surfaces.   This however does not make the system magnetic, because along with \emph{k}, there is  also the opposite vector $\boldsymbol{-k}$ for which the in-plane spin polarization switches the sign, in accordance with time-reversal symmetry [Fig.~\ref{fig:fig2}(d,e)]. When integrated over all spin states and over the entire BZ the net magnetic moment vanishes.

\begin{figure}[tpbh] \centering
\includegraphics[width=0.5\textwidth]{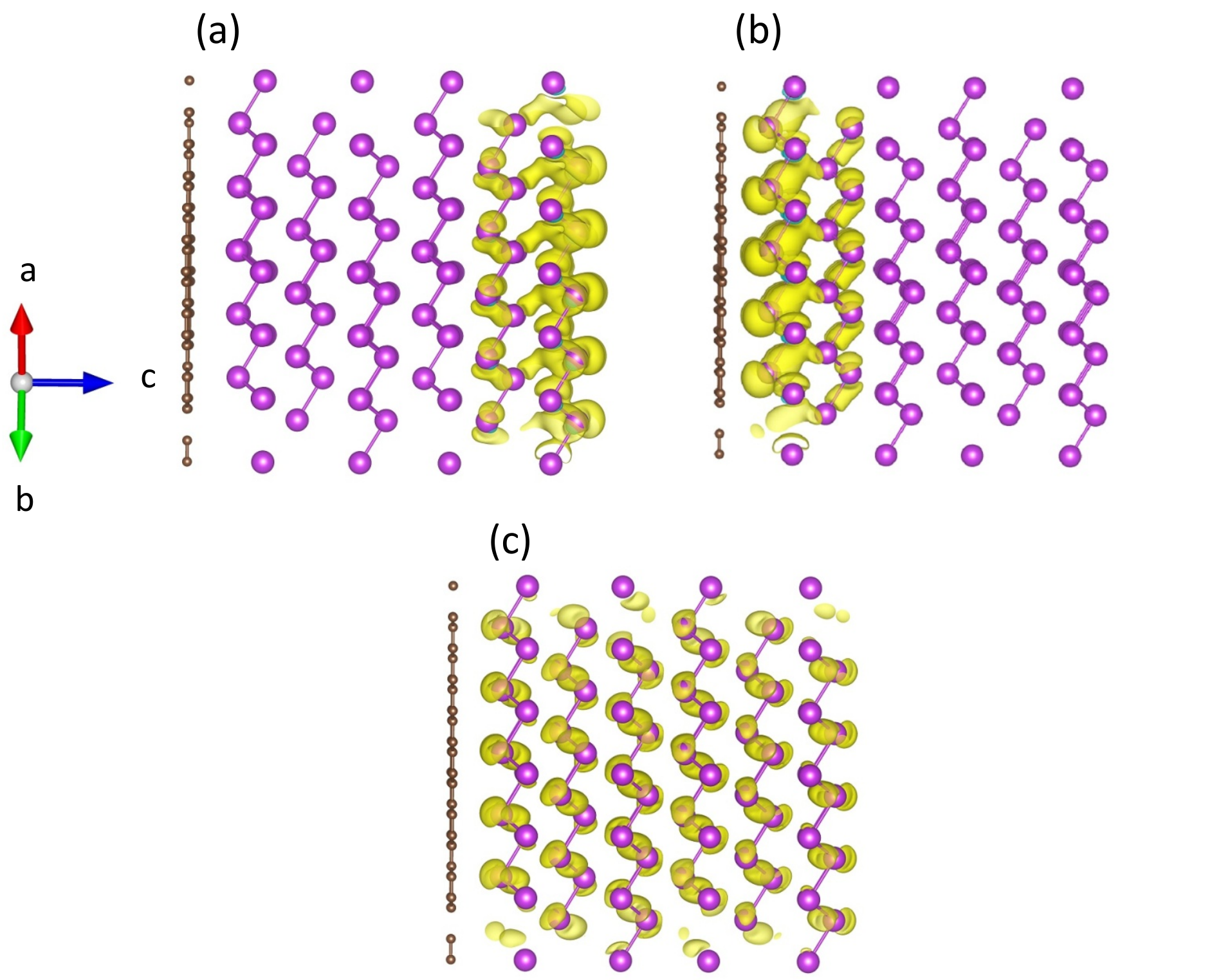}
\caption{Charge density plots for the SS2 surfaces states at two different \emph{k}-points, 1/5($\overline{\Gamma}$ --$\overline{M}$) and $\overline{M}$, which are indicated in Fig.~\ref{fig:fig2}(b) by red circles. Isosurfaces plotted for (a) the lower and (b) the upper (in energy) bands at 1/5($\overline{\Gamma}$ --$\overline{M}$)-point. This \emph{k}-point is simply chosen because at this point, the SS2-states are non-degenerate and split into 2 bands. (c) Charge density at the $\overline{M}$-point, at which the SS2 surfaces states are degenerate. The two bands are now characterized by the same charge distribution.}.

 \label{fig:fig3}
\end{figure}

In the \emph{k}-region where SS1 and SS2 are spin-split, two bands (one from a SS1 pair and one from a SS2 pair) are localized on the graphene/Bi interface and the remaining two on the Bi/vacuum interface. The bands in each pair have opposite spin directions and therefore can be considered as a result of Rashba-type spin-orbit splitting. In the case of ideal Rashba scenario these two bands would cross at the  $\overline{\Gamma}$  and $\overline{M}$ points, but this does not happen here due to quantum size effects.  The bands do not cross at $\overline{\Gamma}$ merely because the used substrate is not thick enough (6 BL). As shown in Ref.~\cite{peng}, the gap between SS1 and SS2 at the $\overline{\Gamma}$ point around E$_F$ disappears for the thin films thicker than 9 BL. The situation is more subtle in the event  of $\overline{M}$ point where the absence of the energy gap  stems from the  strong hybridization between the surface states localized on the opposite sides of the Bi film \cite{koroteev,takayama2, hirahara3, ito1,chang}. In going from $\overline{\Gamma}$ to $\overline{M}$,  the character of SS1 and SS2  gradually changes from surface-like to bulk- or quantum-well-state-like, with the wave functions spreading over the entire thickness of the Bi film. To illustrate this we have explicitly plotted the charge density distributions [Figs.~\ref{fig:fig3}(a,b)] for the SS2 states at the k-points indicated by red circles in Fig.~\ref{fig:fig2}(b). As seen from Figs.~\ref{fig:fig3}(a,b), the spin-split SS2 states at one-fifth of the way from $\overline{\Gamma}$ to $\overline{M}$ are well localized at the Bi/graphene and  Bi/vacuum interfaces. At the $\overline{M}$ point, however, the charge density is uniformly distributed in the region spanning between the two interfaces [Fig.~\ref{fig:fig3}(c)].

In contrast to the $\overline{\Gamma}$ and $\overline{M}$ points, the electronic  states  near the $\overline{K}$ valley are associated with graphene. In the vicinity of the $\overline{K}$-point we clearly observe the graphene-derived Dirac cones. The Dirac cones are separated by a gap of 1.6\,meV [Fig.~\ref{fig:fig2}(c)], which is  much lower than that in structure-I, but still large as compared to 0.024\,meV in pristine graphene (due to SOC)~\cite{abdelouahed,gmitra}. The Dirac points are shifted downwards relative the Fermi level by $\sim$ 0.13\,eV, resulting in \textit{n}-doped graphene, like in structure-I. The degree of doping here, however, is considerably lower than that in structure-I; the reason will be discussed in Section~\ref{doping}.
 
\begin{figure*}[tpbh] \centering
\includegraphics[width=0.65\textwidth]{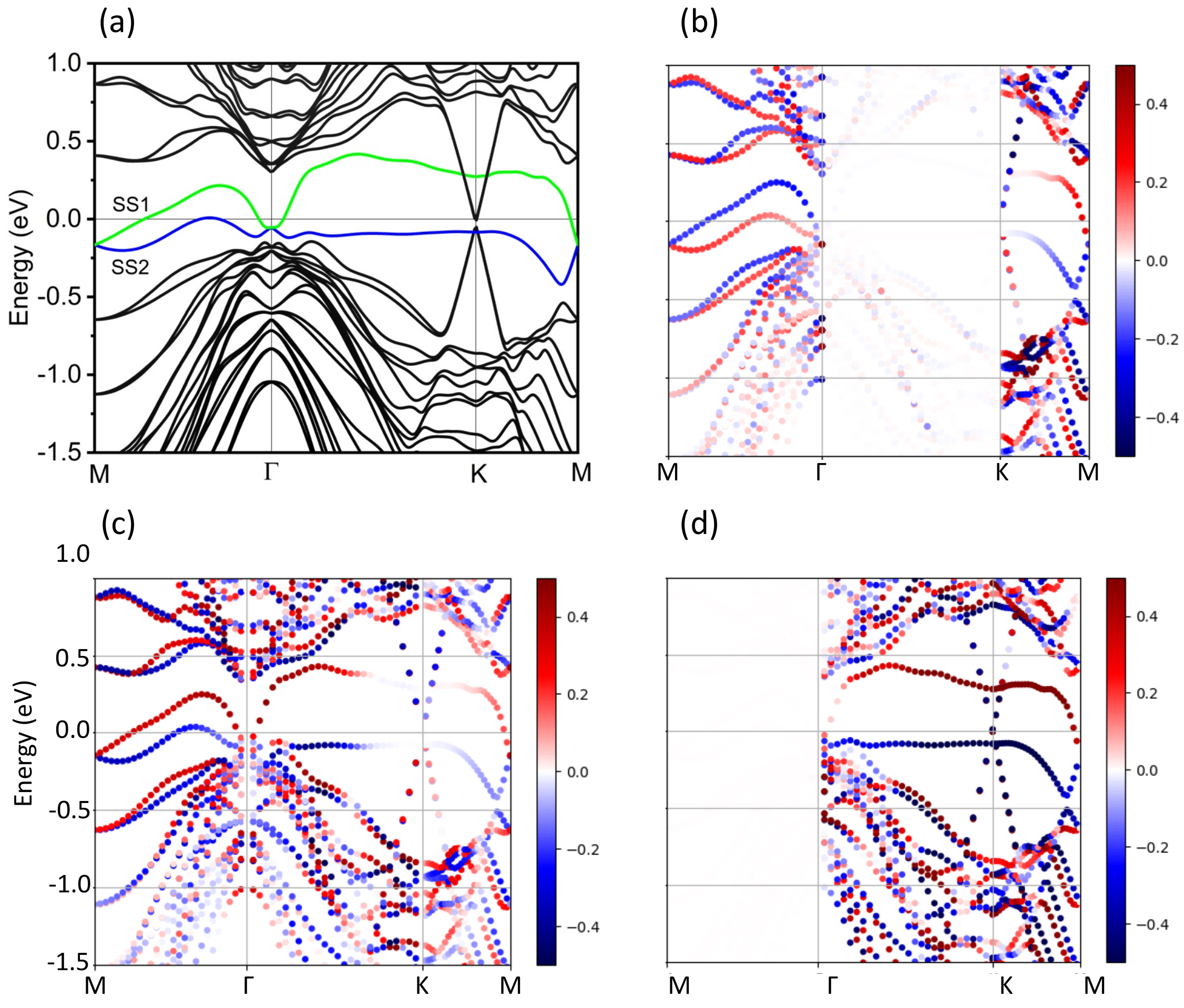}
\caption{Spin-resolved band structure of the  $( 2\times 2)$ Gr on  $(1\times 1)$ Bi (111) heterostrucure with bottom surface terminated with H atoms. (a) With total spin. (b,c and d ) With the resolved  $\pm S_x$,
$\pm S_y$  and  $\pm S_z$  spin  projections, respectively. Colors code the expectation  values  of the projections.}.
 \label{fig:fig4}
\end{figure*}
The other important difference between pristine graphene and graphene on Bi is that in the former the bands are doubly degenerate due to time-reversal and inversion symmetries, whereas in the latter they are spin-split, as seen from the spin-resolved band structure near the $K$ ($K'$) point [Fig. ~\ref{fig:fig2}(c)]. The magnitude of the spin-orbit splitting of the $\pi$ bands is about 1\,meV, and  the spin (its in-plane component) rotates clockwise around the outer Dirac cones (both upper and lower cones). The spin rotates in the opposite sense if one goes from the outer to inner Dirac cone. As to the $z$-component (not shown here), it keeps its sign in going around any cone, but the sign becomes opposite in passing from the upper to lower cone, and vice versa.  All the rotations switch their senses in going from $K$ to $K'$. 

Interestingly, our spin structure is qualitatively similar to that obtained for graphene under out-of-plane electric field, with SOC taken into account~\cite{abdelouahed}. This implies that the Bi substrate affects the electronic structure of graphene as an effective electric field. Since the Dirac bands do not get inverted, as opposed to the case of graphene on WSe$_2$ \cite{gmitra1},  graphene within  this particular heterostructure does not exhibit quantum spin Hall effect.   

\begin{figure*}[tpbh] \centering
\includegraphics[width=0.65\textwidth]{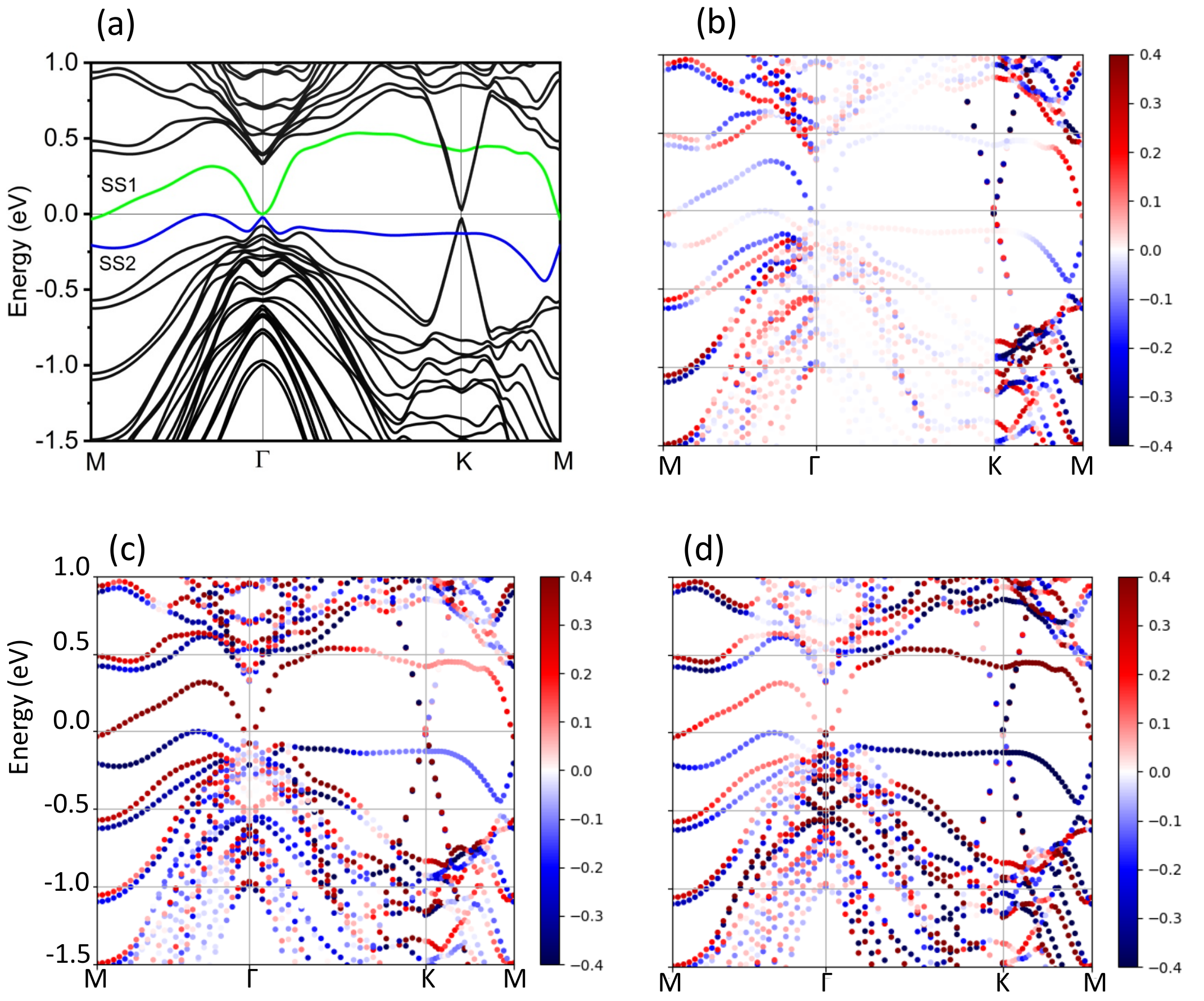}
\caption{Magnetic solution for the  $( 2\times 2)$ Gr on  $(1\times 1)$ Bi (111) heterostrucure with bottom surface terminated with H atoms. (a) The band structure describes the system with a total magnetic moment $\boldsymbol{M}$= [0.03, 0.41, 0.47] $\mu_B$, showing splitting at the TRIM points due to the broken time-reversal symmetry. Band structure plots with the resolved (a) $\pm S_x$, (b) $\pm S_y$, and (c) $\pm S_z$ spin projections, respectively. Colors code the expectation values of the projections.}
 \label{fig:fig5}
\end{figure*}

Strictly speaking, the band structures of the supercells under discussion depend on the relative horizontal shifting of the graphene and Bi lattices, even though this dependence is only slight. One of the visible effects of such a shifting is the displacement of the Dirac cones and changes in their energy gap. In the case of heterostructure shown in Fig.~\ref{fig:fig2}, the Dirac cones (their tips) are located not at the $\overline{K}$ point, but are shifted from it a bit toward $\overline{\Gamma}$ [Fig.~\ref{fig:fig2}(c)]. By assuming that the Dirac cones are centered exactly at $\overline{K}$, one could obtain a wrong energy gap of 11\,meV instead of the actual value of 1.6\,meV.  Further, we found that the gap practically does not change upon "turning off" of SOC. Once again, this confirms the conclusion that the band gap at the Dirac points in the heterostructures under consideration opens up mainly due to fact that the  A- and B-sublattice atoms of graphene ''feel" different potentials induced by the  substrate. 

To mimic the semi-infinite behavior of the Bi (111) film and eliminate interaction between the surface states from the opposite sides, we terminated the  bottom surface with hydrogen. Via structural relaxation, each hydrogen atom was found to sit directly below a Bi atom of the bottom layer, with a bond length of 1.87\,{\AA}.  The introduction of H-atoms removes the surface states SS1 and SS2 from the bottom side, but keeps them on the top side adjacent to the graphene layer. They are spin-split and this splitting is consistent with the Rashba picture [Fig.~\ref{fig:fig4}(a)]. As expected in the Rashba picture, the bands are degenerate at the $\overline{M}$ point. Further, in Fig.~\ref{fig:fig4}(a), one can observe two new bands along the $\overline{\Gamma}$-$\overline{K}$  and $\overline{K}$-$\overline{M}$ lines that cross the upper and lower Dirac cones above and below the Fermi level. They are dominated by H $s$-orbitals hybridized with Bi-$s$ orbitals (in the bottom surface layer closest to H)  and  Bi $p_{x,y}$ orbitals (in the  second bottom layer). The spin texture of the bands is characterized by three main features [Fig.~\ref{fig:fig4}(b-d)]. First, the $x$-component almost vanishes along $\overline{\Gamma}$-$\overline{K}$. This is explained by the fact that in our calculations $\overline{\Gamma}$-$\overline{K}$ is parallel to $x$, whereas the spin orientation tends to be is perpendicular to the wavevector. Second, the out-of-plane ($z$) component is as large as in-plane components, in contrast to the conventional Rashba picture.  And third, the  $z$-component vanishes along the $\overline{\Gamma}$-$\overline{M}$ line. This result can be understood as due to the mirror symmetry of the plane formed by this line and $z$ axis (i.e, the $yz$-plane). Since spin $\boldsymbol{S}$ is an axial vector, its components transform under mirror reflection as $\boldsymbol{S_\perp} \to \boldsymbol{S_\perp}$  and $\boldsymbol{S_\parallel} \to -\boldsymbol{S_\parallel}$, where $\boldsymbol{S_\perp}$ and $\boldsymbol{S_\parallel}$ are the components perpendicular and parallel to the mirror plane. See the detailed discussion by Henk \textit{et al.}~\cite{henk} for a geometrically similar situation of surface states of Au (111). We should note that these three features have been observed in an experiment using photoemission spectroscopy to study Bi(111) thin film grown on Si(111)~\cite{takayama1}. 


Interestingly, in the case of a hydrogenated heterostructure, along with nonmagnetic structure, we also found a magnetic  solution, which is only slightly higher in energy (by several meV).  This solution corresponds to non-collinear magnetism which is maintained by the Bi-atoms in the top surface layer on the graphene side and in the second layer on the bottom (hydrogenated side). The local moments of the Bi-atoms of the topmost surface (graphene side) are mostly parallel to the surface, while the moments on the Bi-atoms from the bottom surface are perpendicular to it (i.e. along $z$). The magnitude of the local moments is calculated to be $\sim$0.10-0.15\,$\mu_B$ for both groups.  

We believe that the surface magnetization in our system has the same origin as the local magnetization in the armchair-edged  BiSb nanoribbons~\cite{lv}. In the case of BiSb nanoribbons, magnetism appears whenever the Sb-atoms along the edges are passivated, but there are unsaturated Bi atoms with dangling bonds. In our case, the Bi atoms from the bottom second layer  also become unsaturated due to passivation of the Bi atoms of the first layer. In addition, the Bi atoms on the topmost surface layer are also effectively unsaturated due to charge transfer to the graphene monolayer (the topmost Bi atoms contribute most to the charge transfer).     

The total energy only weakly depends on the orientations of the local moments, so that it is difficult to find the most favorable orientation of the net magnetic moment of the system, $\boldsymbol{M}$. Fig.~\ref{fig:fig5} shows the calculated band structure for a particular case when  $\boldsymbol{M}$= [0.03, 0.41, 0.47] $\mu_B$. By comparing Figs.~\ref{fig:fig4} and  ~\ref{fig:fig5} one can see that the introduction of magnetic ordering changes the spin texture and the connections of the bands  dramatically due to breaking of the time-reversal symmetry. In particular, as the SS1 and SS2 surface bands disperse along $\overline{\Gamma}$-$\overline{M}$, the $x$-component of their spin switches sign somewhere inside the interval. The bands themselves are not degenerate at $\overline{M}$ since this point is not a surface TRIM any longer. This result sheds light on the heated debate in literature on whether or not SS1 and SS2 in Bi (111) films are degenerate at  $\overline{M}$ [see, for example, work by Chang \textit{et al.}~\cite{chang}]. Some researchers argue that SS1 and SS2 do not converge even in the bulk limit, and therefore, bulk Bi has a nontrivial topology~\cite{ohtsubo, hirahara3}. Our calculations suggest that splitting of the bands at $\overline{M}$ on the (111)-surface of single crystals can be understood by taking into account weak surface magnetism.  This statement is backed by experimental results showing the asymmetry (with respect to $\boldsymbol{k} \to \boldsymbol{-k}$) of spin polarization in Bi (111) on Si (111)~\cite{takayama1, takayama2}.

\subsection{ Structures III and IV: long-period structures}

 \begin{figure*}[tpbh] \centering
\includegraphics[width=0.8\textwidth]{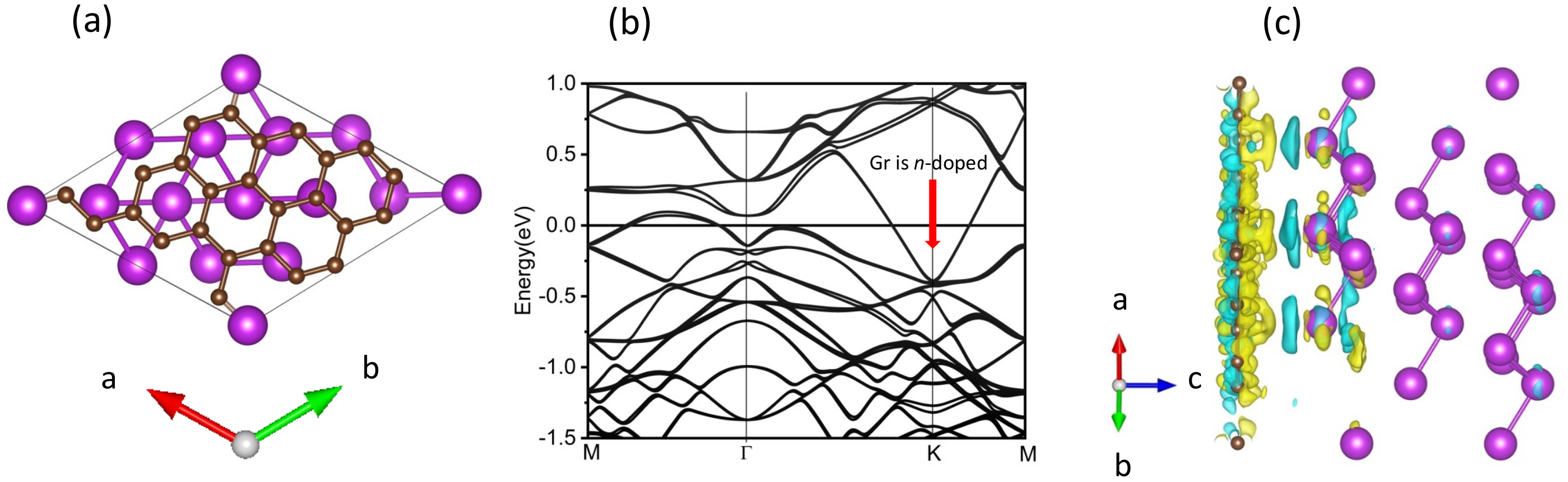}
\caption{(a,b)  A top view of the moir\'e supercell  $(\sqrt{13} \times \sqrt{13})$. The solid black lines enclose an in-plane unit cell of the composite structure. (b) The  band structures for the supersell. The red arrow show the shifting of the the Dirac cones relative to E$_{F}$. (c) The  charge density  $\rho$(Gr/Bi)- $\rho$(Gr)-$\rho$(Bi) showing the redistribution of electrons during the formation of the heterostructure (isosurface at $\pm$ 0.00015). Blue color indicates the depletion, whereas yellow-- accumulation of the electrons.}.
 \label{fig:fig6}
\end{figure*}

 \begin{figure*}[tpbh] \centering
\includegraphics[width=0.8\textwidth]{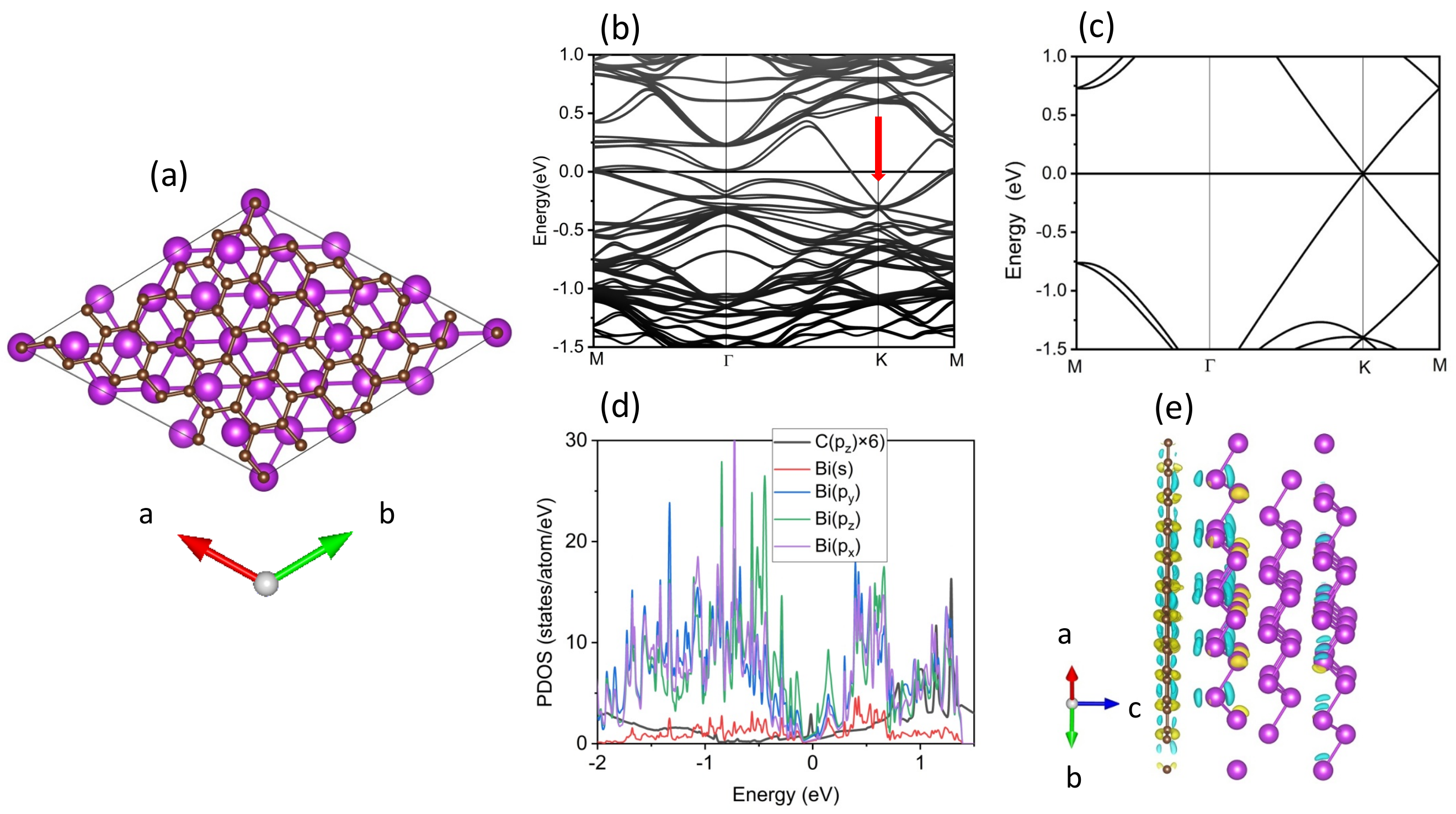}
\caption{(a)  A top view of the moir\'e supercell  $(\sqrt{31} \times \sqrt{31}) $  Gr on $(3\times 3)$ Bi . The solid black lines enclose an in-plane unit cell of the composite structure (b) The  band structure for the superlattice. The red arrows show the shifting of the the Dirac cones relative to E$_{F}$. (c)The  band structure  for a system in which all  Bi are removed while  all the C atoms are kept  in  their positions. (d) The  ``formation" charge density  $\rho$(Gr/Bi)-$\rho$(Gr)-$\rho$(Bi) (isosurface at $\pm 0.0003\,e/\AA^{3}$). }.
 \label{fig:fig7}
\end{figure*}

In these relatively long-period structures [III and IV from Table~\ref{tab:table1}], there are 26 and 62 carbon atoms per supercell, respectively [see Figs.~\ref{fig:fig6}(a) and ~\ref{fig:fig7}(a)]. The characteristic high-symmetry \emph{k}-points of graphene and Bi  map onto different  moir\'e  $k$-points: $\boldsymbol{K}_g \to \boldsymbol{K},\boldsymbol{K}^\prime_g \to \boldsymbol{K}^\prime $ and $\boldsymbol{M}_{Bi}  \to {\Gamma}$ for structure-III, while for structure-IV, we have $\boldsymbol{K}_g \to \boldsymbol{K} ^\prime $, $\boldsymbol{K}^\prime_g \to \boldsymbol{K}$ and $\boldsymbol{M}_{Bi}  \to \boldsymbol{M}$. These foldings allow one to understand the main features of the band structures shown in  Figs.~\ref{fig:fig6}(b) and \ref{fig:fig7}(b)). First, the graphene-derived Dirac cones are well preserved as they are separated from Bi bands in \emph{k}-space, similar to the case of structure-II. Second, in  structure  IV,  the bands close to  E$_{F}$ and along $\overline{\Gamma}$-$\overline{M}$ resemble those in structures I and II -- in all of these cases the $\boldsymbol{M}_{Bi}$-point is translated to $\boldsymbol{M}$.  

In both the band structures, we can clearly see only the upper part of the Dirac cone. The Dirac points are shifted downwards with respect to the Fermi level, by  0.45\,eV and 0.29\,eV for the  structures III and IV, respectively.  As in the previous cases, graphene is \textit{n}-doped, whereas the Bi substrate is \textit{p}-doped. 
To understand the role of different Bi layers in charge transfer, we computed the ``formation charge", or the charge density difference: $\Delta\rho=\rho(Gr/Bi)- \rho(Gr)-\rho(Bi)$, where  $\rho(Gr/Bi)$  is  the charge distribution  in the heterostructure, 
  $\rho(Gr)$ and $\rho(Bi)$ are the distributions in graphene and Bi with the same atomic positions as in the heterostructure [Figs.~\ref{fig:fig6}(c) and ~\ref{fig:fig7}(e)]. One can see that there is charge depletion from the Bi atoms in the top surface layer, closest to the  graphene. Though the  graphene layer accumulates a net charge,  it exhibits stripes of charge  accumulation and depletion periodically arranged in space. Whereas the former usually run along the armchair directions, the latter- along the zigzag directions.  

From the band structures in Figs.~\ref{fig:fig6} and \ref{fig:fig7} one can naively think that the lower part of the Dirac cone (especially for structure-IV) is strongly hybridized with the bands contributed from Bi atoms.
This point also seems to be supported by the comparison of Figs.~\ref{fig:fig7}(b)  and \ref{fig:fig7}(c), showing that in the heterostructure, the  Dirac cone and some of the  Bi-derived bands tend to avoid crossing. 
In reality, however, the hybridization is weak, as seen from the direct calculations of projected density of states, Fig.~\ref{fig:fig7}(d). The situation here is very similar to that found for the structure-I.

  \begin{figure*}[tpbh] \centering
\includegraphics[width=0.9\textwidth]{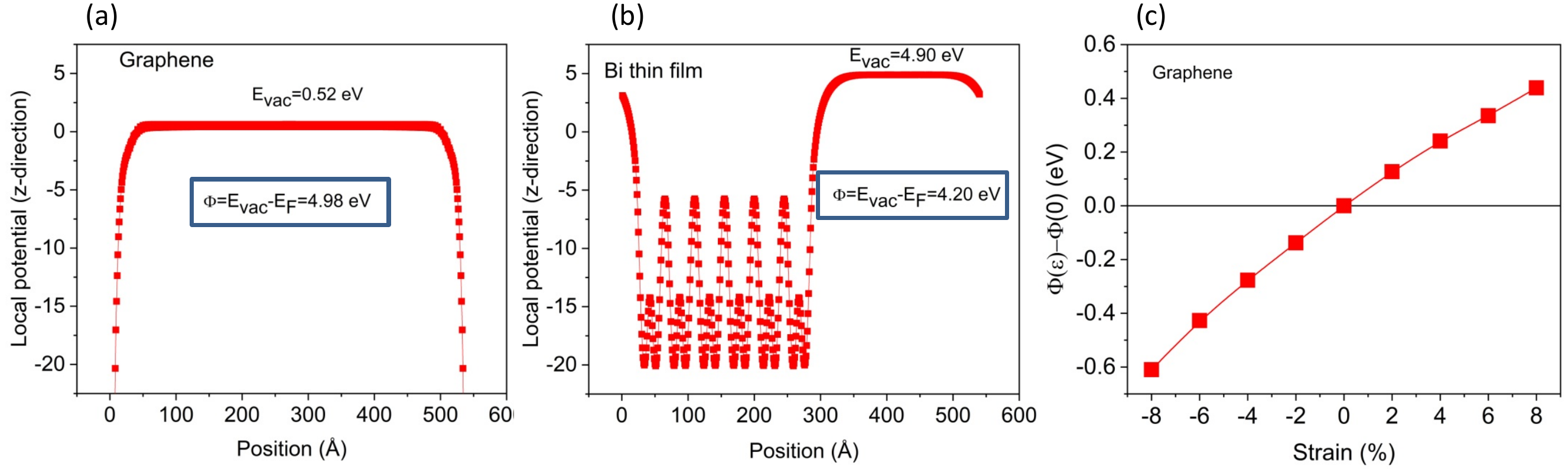}
 \vspace{-12pt}
\caption{(a,b) Local potentials averaged parallel to the (001) surface  as a function of the coordinate z perpendicular to the surface, for graphene and  6-BL-thick Bi(111) thin film,  respectively.  The minima of the potentials correspond to the positions of the atomic layers.  (c) Calculated work function of graphene as a function of strain defined as $(a-a{_0})/a$ }.
 \label{fig:fig8}
\end{figure*}

\section{Charge transfer between graphene and Bismuth} \label{doping}

In order to understand why the level of doping is considerably different for the composite structures I-IV, we calculated the work functions of the constituents -- graphene layer and Bi (111) film.  The work function for a particular surface is defined by the difference between the the local potential in the vacuum and the Fermi level. 
The local potentials averaged parallel to the (001)-surface as a function of the coordinate $z$ for stress-free graphene and Bi (111) film are shown in Figs.~\ref{fig:fig8} (a,b).  
The calculated work function in stress-free graphene is much larger than  in Bi (111) (4.98\,eV vs 4.20\,eV).  Therefore, in a stress-free heterostructure electrons should flow from Bi to graphene, creating an interfacial electric field.

In order to understand why the level of doping itself changes from one structure to another, we should keep in mind that the work function of graphene strongly depends on the strain. As seen in Fig.~\ref{fig:fig8}(c), the work function increases (decreases) with lattice dilation (contraction). In structure-I [$(\sqrt{3}\times \sqrt{3})$  on $(1\times 1)$], the graphene layer, as seen from Table~\ref{tab:table1}, is stretched by 6\%,  and its work function noticeably increases, further facilitating the charge transfer from Bi to graphene. It is not surprising, therefore, that  this structure exhibits the highest degree of $n$-doping of the graphene ($\sim$0.55 eV). In contrast, in structure-II or $( 2\times 2)$  on $(1\times 1)$, the graphene sheet  is compressed by 8\%, and its work function reduces practically to the  value of Bi.  Accordingly, the driving force for the  charge transfer is relatively weak and the Dirac point shifts  down relative to  the Fermi level by only 0.13\,eV.  Structure-III is  similar to structure-I in a sense that  the layer is also stretched, although by the much smaller amount of 2\%.  Consequently, structure-III demonstrates the second largest shifting of the Dirac cone (0.45\,eV).  And finally, structure-IV is almost strain-free; here the shift of the Dirac point is 0.29\,eV.

\section{Discussion}

We are now in a position to discuss the reasons for the experimentally-observed photoresponse in single layer graphene/bismuth nanowire heterostructures.  According to Jin \textit{et al.}~\cite{jin}, it is due to the opening of a band gap at the Dirac point as a result of stretching. This conclusion is based on the results of first-principles calculations~\cite{ni} claiming that  a band gap of ~300\,meV opens up in graphene under a 1\% uniaxial tensile strain.  However, this large gap~\cite{ni} was obtained without taking into account the shift in the position of Dirac points away from the initial reciprocal vectors under a strain that comprises a shear component~\cite{farjam}, thereby over-estimating the gap.  
Moreover, in order to open a band gap of that magnitude in graphene, the strain in graphene should be about 25\% \cite{naumov}, which is much larger than those reported in the experiments~\cite{jin}. On the other hand, our calculations show that a graphene layer is always $n$-doped by the Bi-substrate. This is true even in the structures in which it is stretched, which increases its work function, reducing the extent of its $n$-doping (as compared to the stress-free state) by bismuth. This is consistent with the reports that Bi-atom intercalation results in a sizable $n$-doping of graphene grown on different substrates~\cite{warmuth, zhizhin,hsu}.  Hence, the charge should flow from Bi to graphene, generating a considerable electric field. Thus, our results favor the view of Huber \textit{et al.}~\cite{huber} that the  photocurrent generated in graphene/bismuth nanowire heterostructures is mostly of photovoltaic nature. 

In addition to the aforementioned practical aspect of our study, our work also explores interfacial properties of graphene and Bi composites. We obtain the counter-intuitive result showing that Bi only weakly increases intrinsic SOC in graphene, and the Rashba spin-splitting of the Dirac cone is nearly zero ($\sim$1\,meV). The reason can be understood in the following way. In a free-standing graphene layer, the intrinsic SOC that opens a band gap at the K-point comes almost entirely from carbon's $d-$ and $f-$ orbitals, though their contributions to the density of states are negligible~\cite{abdelouahed}. Hence graphene has only a small intrinsic SOC.  In order to significantly increase the SOC in graphene, one should use a substrate with the $d-$ and $f-$ orbitals capable of hybridizing with the $\pi$-states of graphene (which have a larger contribution to DOS). However, Bi does not belong to this group of substrates, despite the fact that Bi is a relatively heavy element (Z$_{Bi}$ =83). The lighter element, Au (Z$_{Au}$= 79) does a much better job. The $d$-bands of Au strongly hybridize with the $\pi$ bands of graphene, thus leading to a giant proximity-induced Rashba effect \cite{zhizhin}.  As regards to the spin splittings of the Dirac cones, our results point to a negligibly small value. This is consistent with existing experiments and theory. The ARPES data collected around the Dirac cone for Gr/Bi/Ir(111) structure did not reveal any clearly observable spin splitting~\cite{warmuth}. Similar data shows that spin splitting of the Dirac cone in the Gr/Bi/Ni(111) system is finite, but very weak ($\leq$ 10\,meV)~\cite{zhizhin}. In the case of Gr/Bi/SiC(0001), first-principles calculations predict a splitting of 4.99\,meV~\cite{hsu}.  All these spin-splittings are of the same order of magnitude as the value found in our work ($\sim$1\,meV).

Surprisingly, the effects of graphene on the electronic-structure properties of the Bi thin-film are more pronounced. The presence of a graphene layer can be viewed as a perturbation that breaks the mirror symmetry about the central plane parallel to the surface of the substrate, and results in an atypical Rashba spin-orbit coupling. In the case of the non-magnetic solution for the moir\'e superlattice, we demonstrated that this graphene-induced spin-orbit coupling leads to spin-splitting of the surface states on opposite sides of the Bi-film, though without a net magnetic moment.  We also found a magnetic solution which was only slightly higher in energy as compared to the non-magnetic structure by only a few meVs. In the magnetic structure, the spin-splitting results in a net magnetic moment, localized on the surface/sub-surface Bi-atoms. This surface magnetism breaks the time-reversal symmetry, lifting degeneracy of the surface states at the TRIM point ($\overline{M}$), which was also observed experimentally. These results provide a possible explanation for the experiments, which have been at variance with theory.  Theory does, indeed, predicts a non-magnetic ground state, and hence, the degeneracy of the surface states at the TRIM point. However, in experiment, the bismuth film can adopt magnetic structure under different conditions, such as presence of strain, defects and adatoms bonding with the surfaces.

In summary, our study reveals several interesting and non-intuitive phenomena related to the moir\'e physics of the graphene-bismuth heterostructure, while also providing a physical picture required to understand and explain the experiments in this field.

\begin{acknowledgments}
This work was supported W. M. Keck Research Foundation grant and NSF Grant number DMR-1752840. We acknowledge the computational support provided by the Extreme Science and Engineering Discovery Environment (XSEDE) under Project PHY180014, which is supported by National Science Foundation grant number ACI-1548562. We also acknowledge VESTA 3 software for three-dimensional visualization of crystals and volumetric data and the use of \emph{PyProcar} to create figures with spin textures~\cite{PyProcar}.
\end{acknowledgments}



\begin{thebibliography}{99}
\bibitem{novoselov2} K. S. Novoselov,  A. Mishchenko, A. Carvalho,  A. H. Castro Neto, 2D materials and van der Waals heterostructures, Science, aac9439 \textbf{353}   (2016).

\bibitem{cao} Y. Cao, V. Fatemi, S. Fang, K. Watanabe, T. Taniguchi, E. Kaxiras, and  P. Jarillo-Herrero,
Unconventional superconductivity in magic-angle graphene superlattices, Nature,  \textbf{556},43 (2018).

\bibitem{ren2} Y. Ren, Z. Qiao, and Q. Niu, Topological phases in two-dimensional materials: a review, Rep. Prog. Phys. \textbf{79}, 066501 (2016).

\bibitem{dyrdal} A. Dyrda\l{} and J. Barna\ifmmode \acute{s}\else \'{s}\fi{}, Current-induced spin polarization and spin-orbit torque in graphene, Phys. Rev. B \textbf{92}, 165404 (2015).

\bibitem{lee2} J. Lee and J. Fabian, Magnetotransport signatures of the proximity exchange and spin-orbit couplings in graphene,  Phys. Rev. B \textbf{94}, 195401 (2016).

\bibitem{yan} L. Yan, Z. Gu, X. Zheng, C. Zhang, X.  Li,  L. Zhao, and Y. Zhao, Elemental bismuth-graphene heterostructures for photocatalysis from ultraviolet to infrared light, ACS Catal.  textbf{7}, 7043 (2017).

\bibitem{jin} L. Jin, Y. Xiao, D. Zhang, H. Zhang, X. Tang, Z. Zhong, and Q. Yang, Giant optical absorption and low dark current characteristics in wrinkled single layer
graphene/bismuth nanorods, Carbon \textbf{127}, 596 (2018).

\bibitem{huber} T. E. Huber, S. D. Johnson,  J. H. Belk,  J. H. Hunt,  and K . Shirvani, Charge transfer and photocurrent in interfacial junctions between bismuth and
graphene, Phys. Rev.  Appl.  \textbf{10}, 044020 (2018).

\bibitem{akturk} O. \"U. Akt\"urk  and M. Tomak, Bismuth doping of graphene, Appl.  Phys. Lett. \textbf{96}, 081914 (2010).

\bibitem{hsu} C.-H. Hsu, V. Ozolins, F.-C. Chuang, First-principles study of Bi and Sb intercalated graphene on SiC(0001)
substrate. Surf. Sci.  \textbf{616} 149-154, 2013).

\bibitem{warmuth} J. Warmuth, A. Bruix,  M. Michiardi, T. H\"anke, M. Bianchi, J. Wiebe, R.  Wiesendanger, B.  Hammer, Ph. Hofmann, and A. A. Khajetoorians,
Band-gap engineering by Bi intercalation of graphene on Ir(111),  Phys. Rev.  B \textbf{93}, 165437 (2016).

\bibitem{zhizhin} E.V. Zhizhin, A. Varykhalov, A. G. Rybkin, A. A. Rybkina, D. A. Pudikov, D. Marchenko, J. S\'anchez-Barriga, I.I. Klimovskikh, G.G. Vladimirov, O. Rader, A.M. Shikin, Spin splitting of Dirac fermions in graphene on Ni intercalated with alloy of Bi and Au, Carbon,\textbf{93}, 984-996 (2015).

\bibitem{hofmann}Ph. Hofmann, The surfaces of bismuth: Structural and electronic properties, Surf. Sci.,  \textbf{81}, 191-245 (2006).

\bibitem{takayama1} A. Takayama,  T. Sato, S. Souma, and T. Takahashi, Giant out-of-plane spin component and the asymmetry of spin polarization
in surface Rashba states of bismuth thin film,  Phys. Rev. Lett. \textbf{106}, 166401(2011).

\bibitem {lv}H. Y. Lv, H. J. Liu, X. J. Tan, L. Pan,  Y. W. Wen, J. Shi.  and X. F. Tang, The properties of BiSb nanoribbons from first-principles calculations, Nanoscale,   \textbf{4}, 511-517 (2012).

\bibitem{takayama2} A. Takayama, T. Sato, S. Souma, T. Oguchi, and T. Takahashi, Tunable spin polarization in bismuth ultrathin film on Si(111), Nano Lett. \textbf{12}, 1776 (2012).

\bibitem{kresse1}G. Kresse and D. Joubert, From ultrasoft pseudopotentials to the projector augmented-wave method, Phys. Rev. B \textbf{59}, 1758 (1999).

\bibitem{kresse2}G. Kresse and J. Furthmüller, Efficient iterative schemes for ab initio total-energy calculations using a plane-wave basis set,
 Phys. Rev. B \textbf{54}, 11169 (1996).
 
\bibitem{perdew}J. P. Perdew, K. Burke, and M. Ernzerhof, Generalized Gradient
Approximation Made Simple, Phys. Rev. Lett.  \textbf{77}, 3865 (1996).

\bibitem {tkachenko1} A. Ambrosetti, A. M. Reilly, R. A. DiStasio, and A. Tkatchenko, Long-range correlation energy calculated from
coupled atomic response functions , J. Chem. Phys. \textbf{140}, 018A508 (2014).

\bibitem {abdelouahed} S. Abdelouahed, A. Ernst, J. Henk, I. V. Maznichenko, and I. Mertig,  Spin-split electronic states in graphene: Effects due to lattice deformation, Rashba effect, and adatoms by first principles, Phys.  Rev.  B , \textbf{82}, 125424  (2010).

\bibitem {gmitra}   M. Gmitra, S. Konschuh, C. Ertler, C. Ambrosch-Draxl, and J.
Fabian,  Band-structure topologies of graphene: Spin-orbit coupling effects from first principles,  Phys.  Rev.  B , \textbf{82}, 235431  (2009).
\bibitem {gmitra1} M.   Gmitra,  D.  Lochan, P. H\"ogl, and J. Fabian,  Trivial and inverted Dirac bands and the emergence of quantum spin Hall states in graphene on transition-metal dichalcogenides,  Phys.  Rev.  B , \textbf{93}, 155104  (2016).
\bibitem{koroteev}  Yu. M. Koroteev, G. Bihlmayer, J. E. Gayone,  E. V. Chulkov, S. Bl\"ugel,  P. M. Echenique, and Ph. Hofmann, Strong spin-orbit splitting on Bi surfaces,  Phys. Rev. Lett.  \textbf{93}, 046403 (2004).
\bibitem{ito1}  S. Ito, B. Feng, M. Arita, A. Takayama,  R.-Y. Liu,  T. Someya, W.-C. Chen,  T. Iimori,  H. Namatame, M. Taniguchi,
C.-M. Cheng,  S.-J. Tang,  F. Komori,  K. Kobayashi,  T.-C. Chiang,  and I. Matsuda, Proving nontrivial topology of pure bismuth by quantum confinement,  Phys. Rev. Lett. \textbf{117}, 236402(2016).
\bibitem{ohtsubo} Y. Ohtsubo, L. Perfetti, M. Oliver Goerbig, P. Le F\`evre, F.  Bertran, and A. Taleb-Ibrahimi, Non-trivial surface-band dispersion on Bi(111), New J. Phys.  \textbf{15}, 033041 (2013).
\bibitem{hirahara3}T. Hirahara K. Miyamoto, A. Kimura, Y. Niinuma, G. Bihlmayer, E. V. Chulkov, T. Nagao, I. Matsuda, S. Qiao,K. Shimada, H. Namatame, M. Taniguchi and S. Hasegawa, Origin of the surface-state band-splitting in ultrathin Bi films: from a Rashba effect to a parity effect, New J. Phys.  \textbf{10}, 083038  (2008).
\bibitem{chang} T.-R. Chang, Q. Lu, X. Wang, H. Lin, T. Miller T.-C. Chiang, and G. Bian, Band topology of bismuth quantum films,  Crystals \textbf{9}, 510 (2019).


\bibitem {peng} L. Peng, J.-J. Xian, P. Tang A. Rubio, S.-C. Zhang, W. Zhang, and Y.-S. Fu Visualizing topological edge states of single and double bilayer Bi supported,  on multibilayer Bi(111) films,  Phys. Rev. B \textbf{98}, 245108 (2018).

\bibitem {henk} J. Henk,  M. Hoesch, J Osterwalder, A. Ernst, and P. Pruno, Spin-orbit coupling in the L-point surface states of Au(111): spin-resolved photoemission experiments and first-principles calculations, J. Phys.:Condens. Matter \textbf{16}, 7581-7597 (2004).


\bibitem{ni} Z. H. Ni, T. Yu, Y. H. Lu, Y. Y. Wang, and Y. P. Feng, Z. X. Shen,Uniaxial strain on Graphene: Raman spectroscopy study and band-gap opening, ACS Nano 2008, \textbf{2}, 11, 2301-2305.
\bibitem{farjam} M. Farjam and H. Rafii-Tabar,  Comment on “Band structure engineering of graphene by strain: First-principles calculations”, Phys. Rev. B \textbf{80}, 167401(2009).
\bibitem{naumov} I. I. Naumov and A. M. Bratkovsky, Gap opening in graphene by simple periodic inhomogeneous strain, Phys Rev. B\textbf{84} , 245444 (2011).
\bibitem{PyProcar} Uthpala Herath and Pedram Tavadze and Xu He and Eric Bousquet and Sobhit Singh and Francisco Mu\~{n}oz and Aldo H. Romero, PyProcar: A Python library for electronic structure pre/post-processing, Computer Physics Communications, 107080 (2019). 

\url{http://www.sciencedirect.com/science/article/pii/S0010465519303935}
\end{thebibliography}
\end{document}